# Inductive determination of the optimum tunnel barrier thickness in magnetic tunnelling junction stacks for spin torque memory applications

S. Serrano-Guisan[1*], W. Skowronski[2], J. Wrona[2], N. Liebing[1], M. Czapkiewicz[2], T. Stobiecki[2], G. Reis[3], and H. W. Schumacher[1]

1) Physikalisch-Technische Bundesanstalt, Bundesallee 100, D-38116 Braunschweig, Germany

2) AGH University of Science and Technology, Department of Electronics, Al. Mickiewicza 30, 30-059 Krakow, Poland

3) Bielefeld University, Department of Physics, P.O. Box 100131, 33501 Bielefeld, Germany

* Corresponding author:

E-mail:        santiago.serrano-guisan@ptb.de

phone:        +49 (0)531 592 2439,        fax:      +49 (0)531 592 246939

## Abstract

We use pulsed inductive microwave magnetometry to study the precessional magnetization dynamics of the free layer in CoFeB/MgO/CoFeB based magnetic tunnelling junction stacks with varying MgO barrier thickness. From the field dependence of the precession frequency we are able to derive the uniaxial anisotropy energy and the exchange coupling between the free and the pinned layer. Furthermore the field dependence of the effective damping parameter is derived. Below a certain threshold barrier thickness we observe an increased effective damping for antiparallel orientation of free and pinned layer which would inhibit reversible low current density spin torque magnetization reversal. Such inductive measurements, in combination with wafer probe station based magneto transport experiments, allow a fast determination of the optimum tunnel barrier thickness range for spin torque memory applications in a lithography free process.

# Introduction

Spin transfer torque (ST) [1,2] allows the realization of high density magnetic random access memories (MRAM) based on magnetic tunnel junction (MTJ) cells. In such ST-MRAM devices the cells are programmed by a high density current pulse that initiates spin torque precession of the free layer magnetization [3] and eventually induces magnetization reversal[4]. In the MTJ stack the thickness of the tunneling barrier ($t_{MgO}$) plays an important role as it defines two key parameters for MRAM device applications: the resistance area product RA and the tunneling magneto resistance (TMR) ratio. However, both parameters are usually optimum at different thickness ranges of $t_{MgO}$, and therefore a compromise must be found. In addition also the strength of the coupling $J_{FL}$ between the free and the reference layer depends on $t_{MgO}$ [5,6,7] thereby influencing the reversal properties of the ST-MRAM cells. Furthermore, this coupling could influence the effective damping parameter $\alpha$ of the free layer. $\alpha$ is also important for ST applications as the critical current density $j_C$ for ST magnetization reversal is expected to be directly proportional to $\alpha$ [1,8]. Usually, $j_C$ can only be determined in a time consuming process comprising clean room fabrication of ST nanopillars and subsequent magneto transport experiments on individual devices. In contrast α can be determined by fast inductive characterization of the unpatterned MTJ stacks.

Here, we study the time resolved precessional magnetization dynamic of the free layer in CoFeB/MgO based MTJ stacks by pulse inductive microwave magnetometry (PIMM). From a PIMM measurement at a static magnetic field, we derive the free layer precession frequency $f$ and the effective damping parameter $\alpha$. From the field dependence of $f$ we are able to derive the uniaxial anisotropy energy $K_{FL}$ and the coupling $J_{FL}$ between the free and the reference layer. These inductively derived stack parameters are compared to magneto optical Kerr effect (MOKE) measurements of the same stacks. Furthermore, the ST-MRAM key parameters TMR ratio and resistance area product (RA) are determined by using current-in-plane tunnelling (CIPT) technique[9]. The dependence of the derived parameters on $t_{MgO}$ is discussed with respect to ST-MRAM applications. Below a certain threshold barrier thickness we observe an increase of the effective damping for antiparallel configuration of free and reference layer magnetization. Below this threshold thickness the asymmetric $\alpha$ would also induce an asymmetric $j_C$ thereby inhibiting reversible low current density ST magnetization reversal. Therefore, in combination with wafer probe based determination of RA and TMR, such inductive measurements allow a fast determination of the optimum tunnel barrier thickness range for spin torque memory applications.

# Experimental

We study MTJ stacks with a wedge shaped tunneling barrier as sketched in Fig. 1a. The stacks are deposited in a Singulus TIMARIS cluster tool on Si wafers with a layer sequence: Ta(5)/CuN(50)/Ta(3)/CuN(50)/Ta(3)/PtMn(16)/CoFe(2)/Ru(0.9)/$Co_{40}Fe_{40}B_{20}$(2.3)/ $MgO(t_{MgO})$/$Co_{40}Fe_{40}B_{20}$(2.3)/Ta(10)/CuN(30)/Ru(7) from botton to top. Numbers in parentheses refer to the layer thickness in nm. The MgO thickness is varied in the range of $t_{MgO} = 0.62 \ldots 0.96$ nm without changing the other stack parameters using linear dynamic wedge technology. After sputter deposition, the samples are annealed at 350°C for 2 hours with a 0.5 T magnetic field to define the orientation of the exchange bias. For inductive characterization pieces of 2 mm × 4 mm of lateral dimension and of 5 mm x 5 mm for magneto optical characterization were cut from the wafer. Over the size of the characterized wafer pieces the variation of $t_{MgO}$ can be neglected, as the wedge slope was 0.003 nm/mm, and hence each piece is expected to well represent a given barrier thickness.

The measured dependence of RA and TMR on $t_{MgO}$ is presented in Fig 1b. For thick MgO barriers ($t_{MgO} \geq 0.75$ nm) the TMR ratio is very high (TMR > 150%) and almost thickness independent, indicating a good quality of the tunnel barrier and lack of pinholes, while for thinner barriers ($t_{MgO} \leq 0.71$ nm) it drops significantly, pointing out to possible barrier imperfections. On the other hand, RA increases exponentially with $t_{MgO}$ over a broad thickness range, in concordance with previous reports[10,11].

PIMM measurements were performed at room temperature for all MgO thicknesses. Details of our PIMM measurement technique are reported elsewhere [12]. Single PIMM measurements at a given static field deliver the precession frequency $f$ and the effective damping parameter $\alpha$ [13] of the free layer magnetization. Fig. 2 shows typical time resolved PIMM data for $t_{MgO} = 0.76$ nm at three different static field values along the easy axis direction. The time resolved precession traces (open dots in Fig. 2a-c) can be well fitted by an exponentially damped sinusoid (red lines in Fig. 2a-c) [14], showing that the observed magnetization dynamics is always in the linear regime. Furthermore the best fit to the exponentially damped sinusoid allows the determination of the values of $f$ and $\alpha$. It is clear from experimental data that $f$ varies with the applied static field. This field dependence of the free layer precession frequency is plotted in Fig. 2d (open dots).

To derive the important material parameters from Fig. 2d, we model the precession of the free layer within the MTJ stack in a macro spin model of a coupled trilayer system consisting of a free layer, a reference layer and a pinned layer (see Fig. 1a). We assume a 2.3 nm thick free layer with a saturation magnetization $M_s$, uniaxial anisotropy energy $K_{FL}$, and coupling $J_{FL}$ between free layer and the reference layer. Zero net magnetic moment of reference and pinned layer is assumed due to the antiferromagnetic exchange coupling between both layers.

At these conditions, the total magnetic free energy $E$ of the system can be written as:

$$E = -\mu_0 \cdot M_s H \cdot \cos(\phi - \varphi) - K_{FL} \cos^2 \phi - \frac{J_{FL}}{t_{FL}} \cos \phi \quad (1)$$

where $t_{FL}$ is the free layer thickness, and $\phi$ and $\varphi$ are the azimuthal coordinate of the free layer magnetization and the in-plane orientation of the applied magnetic field with respect to the easy axis. Thus, following Smit and Beljers scheme[15], the dispersion relation derived from Eq. (1) is:

$$f = \frac{\gamma \mu_0}{2\pi} \sqrt{\left[H_{ex}\cos\phi + H_k\cos 2\phi + H\cos(\phi - \varphi)\right] \cdot \left[H_{ex}\cos\phi + H_k\cos^2\phi + H\cos(\phi - \varphi) + M_s\right]} \quad (2)$$

where $H_{ex} = \dfrac{J_{FL}}{t_{FL}\left(\mu_0 M_s\right)}$, $H_k = \dfrac{2K_{FL}}{\mu_0 M_s}$ and $\phi$ is obtained by minimizing Eq. (1).

Therefore, fitting the model to the field dependence of $f$ (red line in Fig. 2d) allows to derive the magnetic parameters: $M_s$, $K_{FL}$, $J_{FL}$ and $\alpha$ of our MTJ stacks. Additionally, a magnetic characterization of samples was carried out along both the easy and hard magnetization axis using MOKE magnetometer. An example of a minor easy axis MOKE loop of the same sample is shown in the inset in Fig. 2d (black dots). In order to fit such MOKE loops a complete trilayer coupled system[16] is considered, even though for a small coupling $J_{FL}$ (thicker barriers), minor MOKE loops can be also fitted by considering Eq. (1).

## Results & discussion

### a.) $M_s$ and $K_{FL}$:

Figure 3 shows the MgO thickness dependence of $J_{FL}$, $K_{FL}$ (Fig. 3a) derived from PIMM and MOKE whereas $\alpha$ (Fig. 3b) derived from PIMM. A constant magnetization saturation of $\mu_0 M_s \sim 1.35$ T is obtained for all samples by both techniques (not plotted). Regarding $K_{FL}$, a mismatch is observed between $K_{FL}$ obtained from PIMM (open squares) and MOKE (full squares). While $K_{FL}$ derived from PIMM is almost independent of thickness, the MOKE data show an increase of $K_{FL}$ with decreasing $t_{MgO}$. This deviation up to a factor of 2 is not well understood, but could be related to the different lateral dimensions of samples used for PIMM and MOKE measurements[17] or to an over simplification of the MTJ system by our model.

It is important to note, however, that this mismatch between both $K_{FL}$ values becomes important for tunnel barrier thickness $t_{MgO} < 0.85$ nm. At this region (see Fig. 3a) the coupling contribution to the dispersion relation (third term of Eq. (1)) is at least 5 times larger than the anisotropy contribution (second term of Eq. (1)), suggesting that magnetization dynamics will be mostly characterized by saturation magnetization $M_s$ and $J_{FL}$ coupling. Therefore, this discrepancy between both $K_{FL}$ values should not question our magnetization dynamics analysis.

**b.) J$_{FL}$ coupling:**

In contrast, a large ferromagnetic coupling $J_{FL}$ is observed for all samples, with a very good concordance between PIMM (open dots) and MOKE (full dots) measurements. For $t_{MgO} \geq 0.71$ nm an exponential decrease of $J_{FL}$ as a function of the tunnel barrier thickness is observed. Such a coupling can not be just understood as a Néel "orange-peel" coupling[18,19] arising from the correlated roughness between the free layer and the reference layer. Indeed, assuming a sinusoidal roughness profile, the "orange-peel" effect predicts a monotonic exponential decrease of $J_{FL}$[19]:

$$J_{FL} = \frac{\pi^2}{\sqrt{2}} \left( \frac{h^2}{\lambda} \right) \cdot \mu_0 M_{FL}^2 \cdot \exp\left( -2\pi\sqrt{2} \cdot t_s \middle/ \lambda \right) \qquad (3)$$

where $h$ and $\lambda$ are the amplitude and the wavelength of the roughness and $t_s$ is the barrier thickness. By fitting $J_{FL}$ to this equation (see red line in Fig. 3a) we derive $h = 1.0 \pm 0.7$ nm and $\lambda = 1.1 \pm 0.1$ nm. These values are far too high to be compatible with the large TMR ratios observed for $t_{MgO} > 0.71$ nm in our high quality MTJ stacks (cp. Fig. 1). Furthermore, it has been reported that for this kind of MTJ stacks, the wavelength is 10 nm $< \lambda < 30$ nm[20].

This inconsistency suggests, therefore, that this large FM coupling may be understood as a signature of an interlayer exchange coupling (IEC). The interlayer exchange coupling is an interfacial exchange interaction that appears when two ferromagnetic layers are separated by a thin spacer as a consequence of spin-dependent reflections at the ferromagnetic/spacer interface[5,6]. Unlike for metallic spacers, where IEC oscillates with the spacer thickness [21] a monotonic non oscillatory exponential decrease of IEC as a function of the barrier thickness is expected for magnetic tunnel junctions [5,6,7] (in good concordance with our data). Previous MOKE measurements in epitaxial Fe/MgO/Fe thin films have however shown an antiferromagnetic (AF) interlayer exchange coupling for the same MgO barrier thickness range studied here (0.5 nm $< t_{MgO} <$ 0.9 nm) [22, 23]. Faure-Vincent et al.[22] were able to ascribe the magnitude and the thickness dependence of this AF coupling in the context of the spin-current Slonczewski's model[6]. By doing similarly and considering $E_F = 2.25$ eV, $\Delta = 2.4$ eV and $m_f = 0.75$ $m_e$ for CoFeB ferromagnets [24], we obtain an antiferromagnetic IEC for our samples, in clear contradiction with our data. This controversy could be ascribed to the limited validity of the spin-current Slonczewski's model, which is only valid for thick tunnel barriers ($2 \cdot k \cdot t >> 1$), as pointed out by Katayama et al.[23]. By a full integration of the IEC over the whole Brillouin zone, they show that the strength of the IEC should be ferromagnetic in epitaxial Fe/MgO/Fe thin films, and not antiferromagnetic as observed by Faure-Vincent et al. Moreover, they showed that their data can only be analyzed under Slonczewski's approximation for $t_{MgO} > 1.2$ nm, far away from the thickness range where the AF coupling was observed. These discrepancies therefore show, that spin-dependent tunnelling transport theory in MTJs [5,6] must be improved by, e.g., considering the presence of impurities and defects in the tunnel barrier.

Zhuralev et al. [25] showed that impurities and/or defects in the tunnel barrier can modify the electronic density of states (DOS) and, with a proper concentration, change the strength of the IEC [26]. Katayama et al. used this assumption in order to ascribe their AF coupling due to the presence of O vacancies in their MgO barriers. The presence of such impurities in the tunnel barrier can be produced by material diffusion from the electrode to the MgO layer during the fabrication process. Recent studies showed that a substantial concentration of $BO_x$ in the MgO layer can be obtained after annealing when MgO is deposited by radio frequency sputtering on CoFeB thin films[27]. The presence of such impurities into the tunnel barrier results in an amorphous Mg-B-O layer (with thickness ranging from 1.1 to 2.1 nm [28]) modifying the electronic properties of the MTJ and the energy band of the tunnel barrier[29]. This distortion induces a ferromagnetic IEC for $t_{MgO}$ = 1.1 nm [29] without substantially modifying the large TMR ratios (~150 – 200 %) of such MTJ stacks[28]. All these results are consistent with our data and seem to support Zhuralev's argument [25] in order to explain the origin of the ferromagnetic IEC in our samples. Recently, Yang[30] et al. studied the influence of the relaxation and the oxidation conditions of epitaxial Fe/MgO/Fe stacks on the IEC. It was found that sufficiently oxidation concentration in the Fe/MgO interface can also induce a ferromagnetic interlayer exchange coupling. However, a further systematic study of this impurity-assisted IEC in the sputtered CoFeB/MgO/CoFeB MTJs under a precise control of the parameter conditions during the growth process of the samples should be carried out in order to assert this explanation.

### c.) Effective damping $\alpha$:

The dependence of the effective damping parameter $\alpha$ on $t_{MgO}$ is shown in Fig. 3b. Three different regimes can be observed: A, B and C. For $t_{MgO}$ > 0.76 nm (region A), no significant change of $\alpha$ is observed. Here, $\alpha$ = 0.016 ± 0.003 which is comparable to our previous values obtained by PIMM measurements in single CoFeB layers of similar thickness[31]. This implies that for $t_{MgO}$ > 0.76 nm the influence of neighbouring layers of the MTJ stack on the free layer magnetization dynamics is negligible, and therefore, the observed effective damping parameter $\alpha$ can be ascribed to the intrinsic Gilbert damping $\alpha_0$. Figure 4a-d shows, for $t_{MgO}$ = 0.88 nm, the static field dependence of the precession frequency $f$ (Fig. 4a), the effective damping $\alpha$ (Fig. 4b) and the calculated free layer (FL), reference layer (RL) and pinned layer (PL) magnetization orientation for static fields along the easy axis (Fig. 4c) and along the hard axis (Fig. 4d). Here, a coupled trilayer model is used in order to derive the orientation of all ferromagnetic layers of the MTJ stack [16] by fitting this model to the measurements. Thus, RL refers to the upper CoFeB layer of the synthetic antiferromagnetic layer while PL refers to the bottom CoFe layer of the synthetic antiferromagnetic which is exchange coupled to the PtMn antiferromagnetic layer (see Fig. 1a). We observe that at this tunnel barrier thickness the magnetization reversal of the MTJ stack is similar to a completely free layer system, where the PL and the RL stay along the easy axis while the FL is reversed. The main consequence of this small coupling $J_{FL}$ is a shift of the magnetization loop and the resonance frequency $f$ due to the extra effective easy axis bias field induced by the interlayer exchange coupling between free and reference layers through the MgO. Therefore, for $t_{\mathrm{MgO}} >$ 0.76 nm the magnetization dynamics of the MTJ can be interpreted just in terms of the free layer magnetization. $\alpha(H)$ is symmetric and shows an enhancement at low fields due to inhomogeneous line broadening [13]. This result is of importance for ST-MRAM applications as the critical current density $j_C$ for ST reversal is directly proportional to the effective damping $\alpha$ of the free layer [1,8]. Hence, any additional dissipation of the ST precession by coupling of the free layer magnetization to neighbouring layers would also increase $j_C$ thereby hampering ST-MRAM applications.

However, for thinner tunnel barriers (0.68 nm $< t_{\mathrm{MgO}} <$ 0.76 nm, region B) $\alpha(H)$ becomes asymmetric and a different effective damping parameter is observed for parallel (P) and antiparallel (AP) configurations, with $\alpha_{AP} > \alpha_{P}$. This difference between both damping parameters increases with decreasing $t_{\mathrm{MgO}}$ until, for barrier thickness below $t_{\mathrm{MgO}} \leq 0.68$ nm (region C), only a damped oscillatory signal could be observed at P configurations.

Figure 4e-h shows, for $t_{\mathrm{MgO}} = 0.71$ nm, the static field dependence of the precession frequency $f$ (Fig. 4e), the effective damping $\alpha$ (Fig. 4f) and the free layer (FL), reference layer

(RL) and pinned layer (PL) magnetization orientation for static fields along the easy axis (Fig. 4g) and along the hard axis (Fig. 4h). We can observe from Fig. 4g that except for the reversal process itself, the angular magnetization configurations of the FL, RL and RL are similar to those with $t_{MgO}$ = 0.88 nm. However, in spite of this similarity, a much larger damping parameter is observed when the free layer magnetization is in the AP state ($\alpha_{AP}$ ~ 0.028 ± 0.004) than for P configurations ($\alpha_P$ ~ 0.015 ± 0.003). This implies that our macrospin model used to derive $f$ and $\alpha$ is not sufficient to provide a complete description of the magnetization dynamics of our system at such barrier thicknesses (region B). Indeed, although the frequency spectra can be well described by this approximation, a further analysis must be done in order to understand this asymmetry on $\alpha$. An intuitive explanation for this could be done in terms of the "orange-peel" effect (see Fig. 5). As mentioned above the Néel dipolar coupling results from film roughness favouring the parallel alignment of both FL and RL magnetizations. This roughness leads to a large fluctuating coupling field at the interface of both ferromagnets which, in turn, will induce a local distribution of the magnetization. For thin MgO barriers, this coupling will be strong enough to freeze the magnetic moments of the free layer close to the interface parallel to the reference layer (see Fig.5). Thus, due to the ferromagnetic nature of the coupling, for a P configuration of the stack, these local moments will be almost parallel to the FL magnetization (Figure 5a), and therefore, the effect of the roughness on magnetization dynamics at first approximation should not be important, thereby being tunnel barrier thickness independent, as observed. However, for AP configurations a strong inhomogeneous magnetostatic field is developed in the FL (dashed area in Figure 5b). At such conditions, almost all magnetic moments of the FL are reversed, being some of them still parallel to the RL. As the way the magnetization relaxes towards equilibrium is very sensitive to the details of the microscopic interactions[32], this large magnetic inhomogeneity will lead to an inhomogeneous dephasing of the precession and hence to an increase of $\alpha$. As the Néel coupling increases when $t_{MgO}$ decreases (see Eq. 5) this effective damping asymmetry will be larger for thinner tunnel barriers, as observed in our experiments.

Finally, for $t_{MgO} \leq 0.68$ nm (region C) magnetization precession is only observed at P configurations. At these thickness range $J_{FL}$ is too strong and of the same order of magnitude as both the antiferromagnetic exchange coupling between the pinned layer and the reference layer ($J_{AF} \sim -221$ μJ/m$^2$) as well as the exchange bias coupling between the pinned layer and the pinning layer ($J_{ex} \sim 188$ μJ/m$^2$) [11]. This implies that all ferromagnetic layers of the MTJ stack are mostly coupled leading in a macrospin model to a scissored state of the SyAF pinned layer when the free layer magnetization is being reversed. Consequently, at these conditions the whole system is involved in precessional magnetization dynamics and PIMM data can no longer be analyzed by Eq. (2). Furthermore, the hysteretic behaviour of the FL magnetization reversal is not observed at such conditions (not shown). This behaviour imposes therefore, a minimum tunnel barrier thickness ($t_{MgO} > 0.68$ nm) that must be considered in order to ensure the existence of a bistable state suitable for MRAM applications.

Note that magneto transport measurements have been previously performed on patterned MTJ nanopillars fabricated from the same MTJ stacks in order to study the thickness dependence of the threshold current density for ST magnetization reversal $j_c$ [11]. Such measurements reveal that $j_c$ shows similar barrier thickness dependence as the effective damping parameter derived from PIMM measurements. Furthermore, they also showed that for thin tunnel barriers ($t_{MgO} < 0.76$ nm) the evolution of the threshold current density for AP-P transitions ($j_c^{AP}$) is higher than for P-AP transitions ($j_c^{P}$). This behaviour can not be ascribed to a different polarization factor on Slonczweski's expression of $j_c$ [8] due to the different orientation of the free layer magnetization with respect to the reference layer. However, this effect could indeed be a consequence of the different damping parameters for P and AP configurations as observed in our measurements. These results imply that, owing to the intimate relationship between $j_C$ and the effective damping parameter $\alpha$, inductive measurements are an excellent tool to investigate the ST-MRAM key parameter $j_C$ without time consuming lithographic processes for patterning MTJ nanopillars and hence derive the optimum tunnel barrier thickness range for efficient ST-MRAM devices.

In order to develop optimum ST-MRAM devices with magnetic memory cells below 100 nm, MTJ stacks must show the following features: i) RA product $< 5\ \Omega \cdot \mu m^2$ (for impedance matching with peripheral circuits), ii) high enough TMR ratios (in order to generate large output signals and highly enough signal-to-noise ratios[33]), and iii) low ST magnetization reversal currents $j_c$, and hence low $\alpha$. In our MTJ stack under investigation such conditions are fulfilled for a tunnel barrier thickness range between $0.76\ \text{nm} \leq t_{\text{MgO}} \leq 0.85\ \text{nm}$ (see Fig. 1b and Fig. 3), in agreement with magneto transport experiments in patterned nanopillars fabricated from the same MTJ stacks[11]. This optimum tunnel barrier thickness range can hence be determined by combination of probe station based magneto transport (i,ii) and inductive measurements (iii) of unpatterned MTJ stacks and thus in a fast and lithography free characterization process.

## Conclusions

In summary, we have characterized sputtered CoFeB/MgO/CoFeB magnetic tunnel junction (MTJ) stacks with different MgO thicknesses ($0.61\ \text{nm} \leq t_{\text{MgO}} \leq 0.96\ \text{nm}$) by Magneto Optical Kerr Effect (MOKE) and by Pulsed Induced Microwave Magnetometry (PIMM) and wafer prober current-in-plane tunnelling (CIPT). From these measurements the precession frequency spectra $f$, the free layer anisotropy field $K_{FL}$, the exchange coupling between the free layer and reference layer $J_{FL}$ and the effective damping parameter $\alpha$ as well as their tunnel barrier thickness dependence were derived. Furthermore the electrical parameters RA and TMR were determined. A large ferromagnetic exponential decrease of $J_{FL}$ with decreasing barrier thickness has been observed which might arise from an impurity-assisted interlayer exchange coupling. By taking into account the thickness dependence of the TMR ratio, of the RA product and $\alpha$, the optimum tunnel barrier thickness range for low current ST-MRAM devices is determined. For our MTJ stack under investigation this optimum thickness range is between 0.76 nm $\leq t_{MgO} \leq$ 0.85 nm. This fast and lithography-free determination of the optimum barrier thickness range is in excellent concordance with studies of the critical current density $j_c$ dependence on $t_{MgO}$ derived by conventional magneto transport experiments on individual patterned nanopillars fabricated from the same MTJ stacks. This proves the potential of inductive characterization as a fast and efficient characterization tool for optimization and testing of ST materials.

## Acknowledgments

We would like to thank J. Langer and B. Ocker from Singulus AG for sample growth and helpful discussions. S.S.G., N.L and H.W.S. acknowledge funding from European community's Seventh Framework Programme, EEA-NET Plus, under IMERA-Plus Project-Grant No. 217257. W.S. and T.S would like to thank the Foundation for Polish Science MPD Programme co-financed by the EU European Regional Development Fund and the Polish Ministry of Science and Higher Education grants (IP 2010037970 and NN 515544538) and SPINSWITCH Project MRTN-CT-2006-035327.

**FIGURES:**

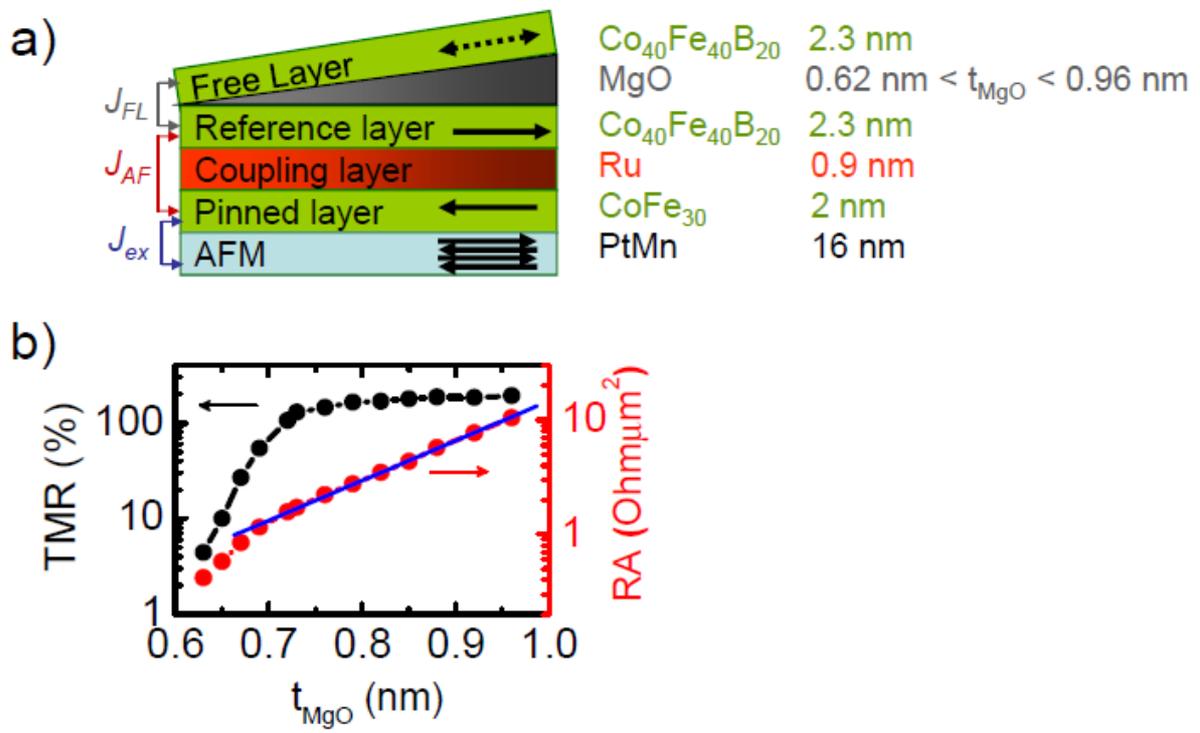

**Figure 1**

S. Serrano-Guisan et al.

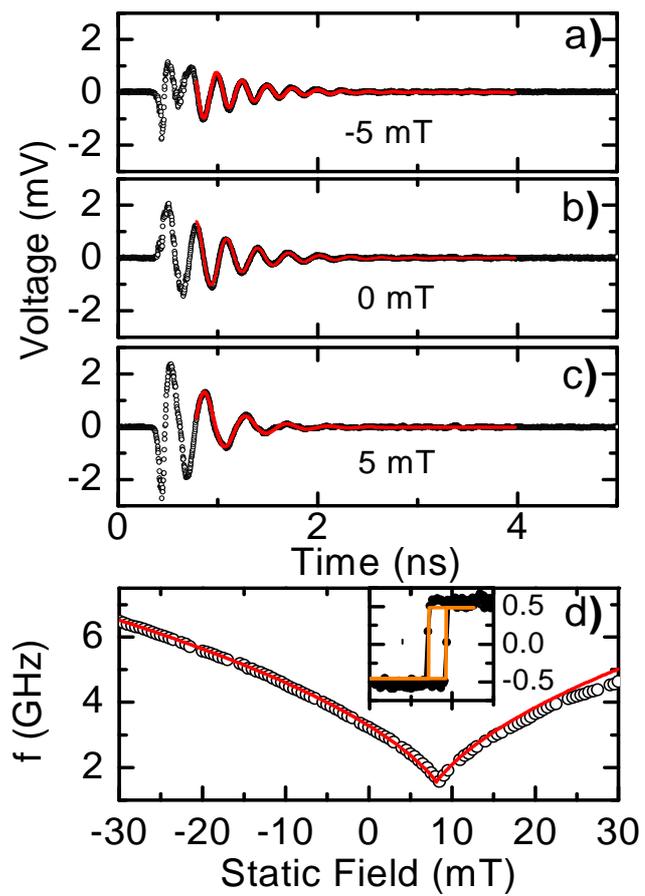

**Figure 2**

**S. Serrano-Guisan et al.**

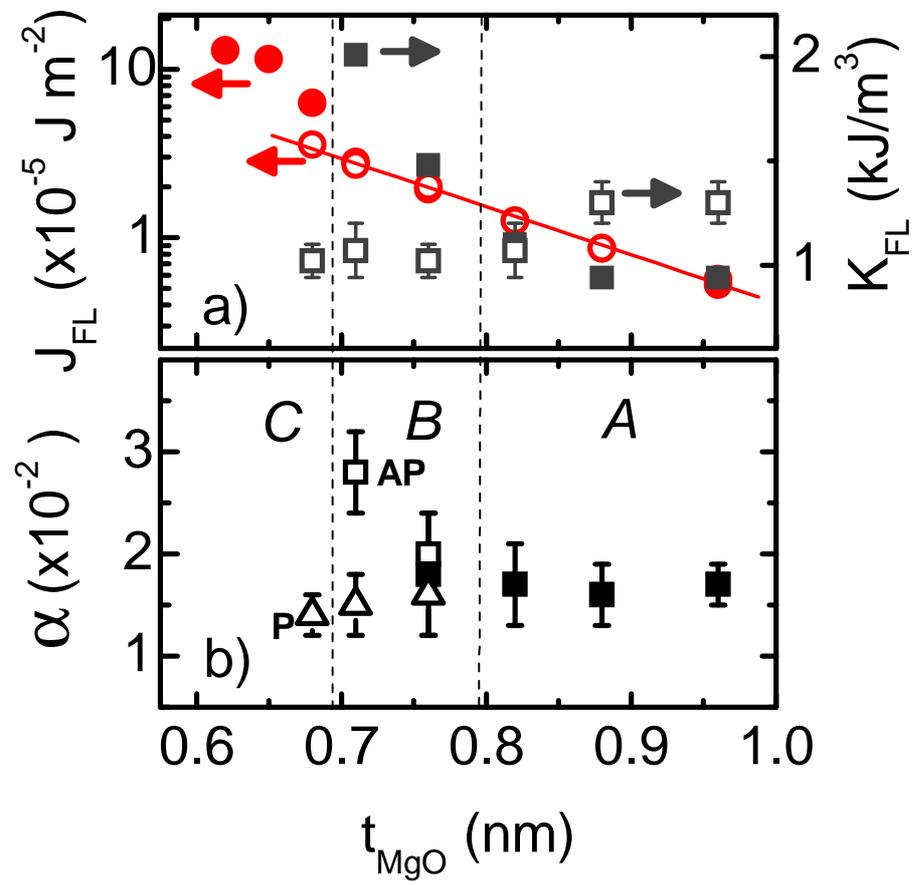

**Figure 3**

**S. Serrano-Guisan et al.**

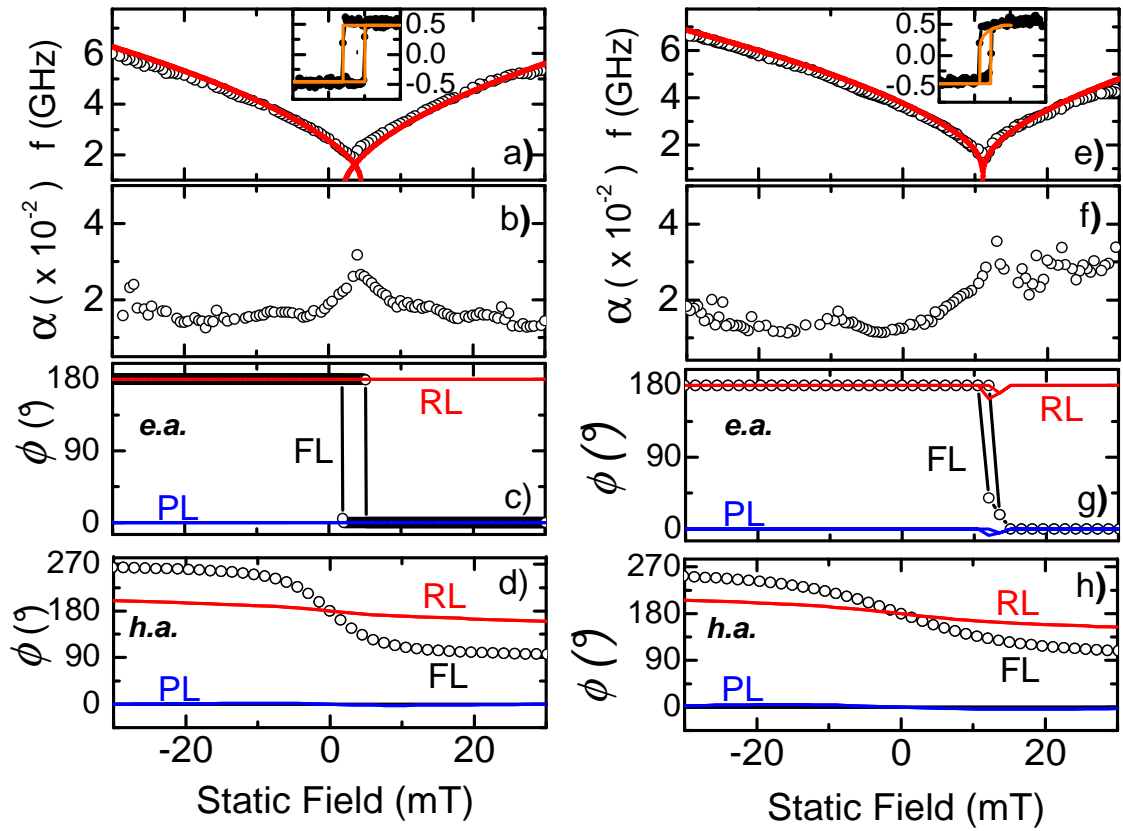

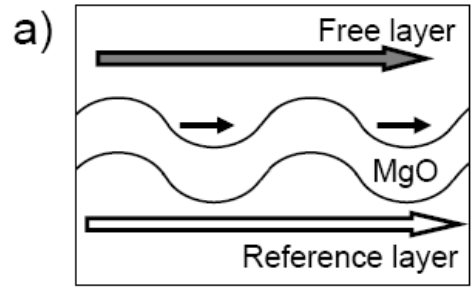

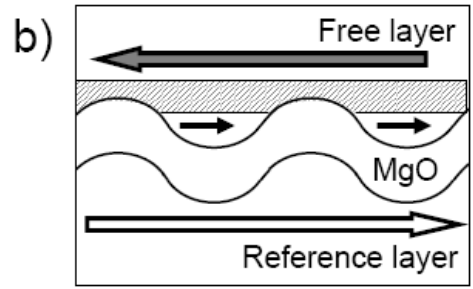

**Figure 5**

**S. Serrano-Guisan et al.**

**FIGURE CAPTIONS:**

**Fig 1:** a) Schematic picture of the MgO based MTJ stack. $J_{FL}$ refers to the exchange coupling between the free layer (FL) and the reference layer (RL), $J_{AF}$ refers to the antiferromagnetic interlayer exchange coupling between the reference layer (RL) and the pinned layer (PL) whereas $J_{ex}$ refers to the exchange bias coupling between the antiferromagnetic pinning layer (AF) and the pinned layer (PL) b) TMR and RA product as a function of MgO barrier thickness measured by CIPT technique (ref.9). Blue line shows the exponential thickness dependence of RA.

**Fig 2:** a)-c) PIMM data (open dots) for $t_{MgO}$ =0.76 nm at different easy axis fields. Red lines show fits by an exponentially damped sinusoid d) Static field dependence of the precession frequency derived from PIMM (open dots). Red line shows the dispersion relation of a Stoner-Wolfarth single-domain model with $H_K$ = 1.8 mT and $J_{FL}$ = 20.1 µJ/m$^2$ .Inset: MOKE loop (black dots) and single domain model approximation (orange line).

**Fig 3:** MgO thickness dependence of a) the exchange coupling $J_{FL}$ between the free layer and the reference layer (circles) as well as the uniaxial anisotropy energy $K_{FL}$ of the free layer (squares) and b) the effective damping parameter. Open (full) symbols in Fig. 3a are referred to values of $J_{FL}$ or $K_{FL}$ derived from PIMM (MOKE) measurements. Red line shows the Néel coupling behaviour derived by fitting $J_{FL}$ data to Eq. (3). In Fig. 3b open triangles refer to the damping parameter derived at parallel configurations ($\alpha_P$), open squares are referred to the damping parameter at antiparallel configurations ($\alpha_{AP}$), and full squares are referred to the averaged damping parameter between both configurations. At MgO thickness $t_{MgO} > 0.76$ nm both configurations have the same effective damping parameter ($\alpha_{AP} = \alpha_P \equiv \alpha$ ).

**Fig 4:** (a), (e) Dispersion relation and minor MOKE loops, (b), (f) effective damping dependence on easy axis magnetic fields and (c-d), (g-h) simulated magnetic field dependence of magnetization orientation of each ferromagnetic layer for (c), (f) easy axis (e.a.) and (d), (h) hard axis (h.a.) magnetic fields with $t_{MgO} = 0.88$ nm ((a), (b), (c) and (d)) and $t_{MgO} = 0.71$ nm ((e), (f), (g) and (h)). (a), (e) Open dots (red line) show the measured (simulated) resonance frequency. Inset: Full dots (orange line) shows the measured (simulated) minor MOKE loops. (c-d), (g-h) FL, RL and PL refer to the free layer, reference layer and pinned layer respectively.

**Fig 5:** Schematic picture of the cross-sectional profile of the MTJ stack. The Néel coupling induces a FM coupling between the reference layer and the magnetic moments located at the "valleys" of the rough free layer. For thin enough MgO barriers the Néel coupling is so strong that those magnetic moments (small black arrows) are always aligned to the reference layer magnetization (white arrow). For pararallel configurations (a) the whole free layer is parallel aligned to the reference layer, whereas for antiparallel configurations (b) just a fraction of the free layer is reversed and a region with a large inhomogeneous distribution of the magnetization is developed (dashed area).

and the antiferromagnetic pinning layer and $K_{AF}$ is the anisotropy constant of the antiferromagnetic pinning layer.